\documentclass[apjl]{emulateapj}
\usepackage{epsfig}
\begin{document}

\title{
Low-velocity streams in the solar neighborhood caused by the Galactic bar
}
\author{I.~Minchev\altaffilmark{1}, C.~Boily\altaffilmark{1}, 
A.~Siebert\altaffilmark{1} and O.~Bienayme\altaffilmark{1}}

\altaffiltext{1}{Universit\'e de Strasbourg, CNRS, Observatoire Astronomique, 11 rue
de l'Universit\'e, 67000 Strasbourg, France; minchev@astro.u-strasbg.fr}

\begin{abstract}
We find that a steady state bar induces transient features at low velocities in 
the solar neighborhood velocity distribution due to the initial response of the 
disc, following the formation of the bar. We associate these velocity streams with two 
quasi-periodic orbital families, librating around the stable $x_1(1)$ and $x_1(2)$ 
orbits near the bar's outer Lindblad resonance (OLR). In a reference frame moving 
with the bar, these otherwise stationary orbits precess on a timescale dependent on 
the strength of the bar, consistent with predictions from a simple Hamiltonian model 
for the resonance. This behavior allows the two orbital families to reach the solar 
neighborhood and manifest themselves as clumps in the u-v plane moving away from 
($x_1(2)$), and toward ($x_1(1)$) the Galactic center. Depending on the bar parameters 
and time since its formation, this model is consistent with the Pleiades and Coma 
Berenices, or Pleiades and Sirius moving groups seen in the Hipparcos stellar velocity 
distribution, if the Milky Way bar angle is $30^\circ\la\phi_0\la45^\circ$ and its 
pattern speed is $\Omega_b/\Omega_0=1.82\pm0.07$, where $\Omega_0$ is the angular 
velocity of the local standard of rest (LSR). Since the process is recurrent, we can 
achieve a good match about every six LSR rotations. However, to be consistent with the 
fraction of stars in the Pleiades, we estimate that the Milky Way bar formed $\sim2$ 
Gyr ago. This model argues against a common dynamical origin for the Hyades and 
Pleiades moving groups.
\end{abstract}


\section{Introduction}

\begin{figure*}
\epsscale{1.2}
\plotone{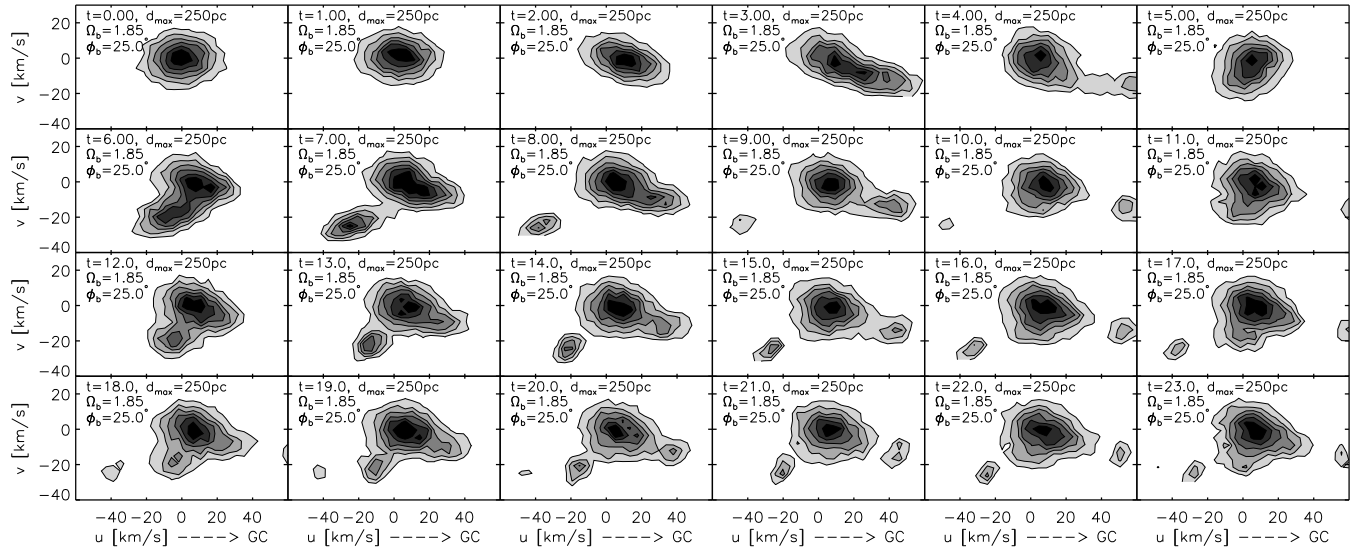}
\figcaption{
Time development of the $u-v$ plane for a simulation of a stellar disc with
Milky Way kinematics perturbed by a steady state bar. The pattern speed is
fixed at $\Omega_b/\Omega_0=1.85$, the maximum sample depth is $d_{max}=250$ pc, 
and the bar orientation is $\phi_0=25^\circ$ (geometry as shown in figure 
\ref{fig:shots}). Shaded contours show the particle number density.
Each panel is a snapshot of a particular time of the same simulation, in increments 
of one rotation at $r_0$. Note that even though the bar perturbation is stationary, 
structure varies periodically with time.
\label{fig:uv185}
}
\end{figure*}

The Milky Way bar is hard to observe directly due to our position in the
Galactic disc. Thus, its parameters, such as orientation of its major axis with
respect to the Sun-Galactocentric line and angular velocity, have been inferred 
indirectly from observations of the inner Galaxy (e.g., 
\citealt{blitz91,weinberg92}). 

However, numerical modeling suggests that the bar can also affect the local 
stellar velocity distribution. It has been found to account for the Hercules 
stream \citep{quillen03,fux01,dehnen00}, the vertex deviation 
\citep{muhlbauer03}, errors in the Oort constants \citep{mnq07}, and (in
combination with spiral structure) the observed flatness in the age-metallicity
relation \citep{mf09}. The bar 
angular velocity, or pattern speed, $\Omega_b$, has been well established to be such 
that the solar circle lies just outside the bar's 2:1 outer Lindblad resonance (OLR). 
\cite{dehnen00} found that in order to reproduce the Hercules stream, the bar
pattern speed should be $\Omega_b/\Omega_0=1.85\pm0.15$, where $\Omega_0$ is 
the angular velocity of the local standard of rest (LSR), and the bar orientation 
with respect to the Sun-Galactocentric line is in the range 
$10^\circ<\phi_0<70^\circ$. By accounting for trends seen on the Oort constant C, 
\cite{mnq07} found $\Omega_b/\Omega_0=1.87\pm0.04$ and $20^\circ<\phi_0<45^\circ$. 
Such bar orientation is consistent with estimates derived from IR photometry 
($15^\circ<\phi_0<45^\circ$) and OGLE-II microlensing observations of red clump 
giants in the Galactic bulge ($24^\circ<\phi_0<27^\circ$, \citealt{rattenbury07}).

Recent analysis of the local velocity field show that the origin of the most
prominent low-velocity moving groups in the solar neighborhood favor a dynamical 
origin \citep{famaey08}. While the bar has been shown to affect the u-v plane at
higher velocities, spiral density waves have been found to create resonant 
structure in the lower velocity regions, such as the splitting of the 
Pleiades/Hyades and Coma Berenices moving groups \citep{qm05}. Is it really true 
that the bar is not effective at low velocities? \cite{fux01} investigated the 
effect of the bar on the local velocity distribution, similarly to the work by 
\cite{dehnen00}, but for test particles integrated forward in time. The author 
reports time varying structure in the u-v plane following the growth of the bar.
However, he did not investigate this further, being more interested in
the effect of the bar once the disc has fully responded to the perturbation.
To increase the particle statistics and reduce phase mixing, the u-v 
distributions presented in the work by \cite{fux01} were averaged over 10 bar
rotations and then smoothed. 

However, it is possible that the effect of the initial response of the disc to
the formation of a central bar can still be 
seen today in the solar neighborhood velocity distribution, provided 
the bar has formed or evolved recently. There are both observational and 
theoretical arguments suggesting such a scenario. By using data from the 
Two Micron All-Sky Survey (2MASS), \cite{cole02} estimated that the Milky 
Way bar is likely to have formed more recently than 3 Gyr ago and suggested 
that this event could have been triggered by a now-merged
satellite. \cite{minchev09a} showed that phase wrapping in the
thick disc caused by initial conditions which might have been left following a 
merger, can explain four observed high velocity streams. To match the locations 
of these streams, a strong perturbation of the MW disc is required 
about 2 Gyr ago. Such an event could have triggered the formation of the MW bar.  
In addition, cosmological simulations show that massive minor mergers are
likely to have happened during the lifetime of Milky Way size system. For 
example, \cite{kazantzidis08} estimated that as many as five objects more 
massive than 20\% $M_{disc}$ could have been accreted since $z\sim1$.  
All this evidence from observations and simulations calls for a detailed
investigation of the effect of a young, or a recently evolved bar, on the 
solar neighborhood velocity distribution. 

In this paper we examine the effect of a central bar on the local velocity
distribution by simulating the time evolution of a barred stellar disc. 
Unlike in previous works, here we do not time average over particles positions,
in order to investigate in detail the response of the disc. We explore the
possibility that stellar streams in the u-v plane could be induced by 
a recent bar formation or evolution.

\begin{figure*}[t!]
\epsscale{1.2}
\plotone{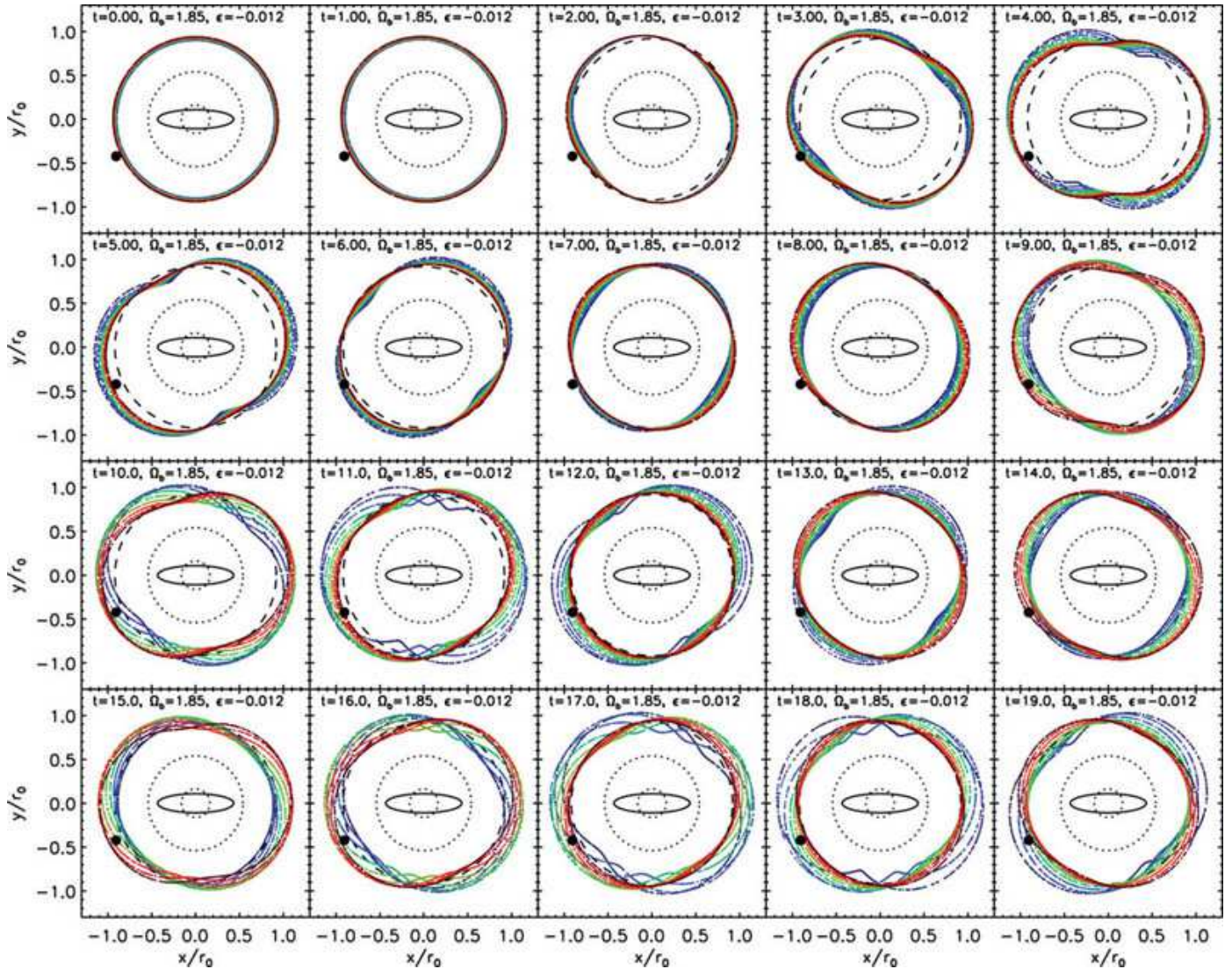}
\figcaption{
Time development of nine, initially circular rings near the bar 2:1 OLR, in the bar
reference frame. All simulation parameters are the same as in figure \ref{fig:uv185}. 
The 2:1 ILR resonance and corotation are shown as dotted circles; the 2:1 OLR 
location is indicated by the dashed circle. The solar neighborhood is shown by a 
black dot, just outside the 2:1 OLR at an angle of $25^\circ$ lagging the bar. 
Two families of orbits associated with the $x_1(1)$ (red) and $x_1(2)$ (blue) 
orbits precess at different rates, giving rise to the streams in figure 
\ref{fig:uv185}. 
\label{fig:shots}
}
\end{figure*}

\section{Simulation set up}

We perform test-particle simulations of a stellar galactic disc with parameters
consistent with Milky Way kinematics.
The density distribution is exponential, $\Sigma(r)\sim e^{-r/r_\rho}$,
with a scale length $r_\rho = 0.38r_0$, consistent with the 2MASS photometry
\citep{ojha01}. 
We give particles an initial radial velocity dispersion in the form of a Gaussian 
distribution. The Milky Way disc is known to have a radial velocity dispersion
which decreases roughly exponentially outwards:
$\sigma^2_u\sim e^{-r/r_{\sigma^2}}$. In accordance with this we implement
an exponential decrease in the standard deviation of the radial velocity
dispersion of stars with radius, with a scale length $r_{\sigma^2}=0.45r_0$
\citep{lewis89}. Since we are interested in the low-velocity regions of the 
u-v plane, we set $\sigma_u=10$ km/s at $r_0$.

To explore the time development of the system, we do not time-average over
position and velocity vectors, as it is frequently done in test-particle 
simulations (e.g., \citealt{fux01,mnq07}) where no dynamical development of 
the system is expected. 
For each simulation we integrate N=$5\times10^6$ particles for 30 rotation 
at the solar radius, $r_0$.

The background axisymmetric potential due to the disc and halo has the form
$\Phi_0(r)=v_0^2\log(r)$, corresponding to a flat rotation curve.
We model the nonaxisymmetric potential perturbation due to the Galactic bar
as a pure quadrupole
\begin{equation}
\Phi_{\rm b} = A_{\rm b}(\epsilon_{\rm b}) \cos[2(\phi-\Omega_{\rm b}
t)]\times\left\{
\begin{array}{cclcr}
         \left(r_{\rm b}\over r\right)^3  &,&  r&\ge & r_{\rm b}    \\ 
       2-\left(r\over r_{\rm b}\right)^3  &,&  r&\le & r_{\rm b}
\end{array}
\right.
\end{equation}
Here $A_{\rm b}(\epsilon_{\rm b})$ is the bar's gravitational potential
amplitude, identical to the same name parameter used by \cite{dehnen00};
the strength is specified by $\epsilon_{\rm b}=-\alpha$
from the same paper.
The bar length is $r_{\rm b}=0.8r_{\rm cr}$ with $r_{\rm cr}$ the bar
corotation radius. The pattern speed, $\Omega_{\rm b}$ is kept constant.
The bar amplitude $\epsilon$ is initially zero, grows linearly with time at 
$0<t<t_1$ and transitions smoothly to a constant value after $t=t_1=4$ bar 
rotations. This insures a smooth transition from the axisymmetric to the 
perturbed state. 

In our units the solar neighborhood (SN) radius is $r_0=1$; the circular speed 
is $v_0=1$ everywhere since the rotational curve is flat. To convert to real 
units we use Local Standard of Rest (LSR) tangential velocity of 240 km/s, and 
Galactocentric distance of 8 kpc. The 2:1 OLR with the bar is achieved when
$\Omega_{\rm b}/\Omega_0=1+\kappa/2\approx1.7$, where $\kappa$ is the epicyclic
frequency. For a flat rotation curve $\kappa=\sqrt{2}\Omega_0$.

\section{Results}

For an axisymmetric disc the potential is independent of time, thus
energy and angular momentum are conserved.
In the case of a single periodic perturbation, which is stationary with time,
there is still a conserved quantity in the reference frame rotating with the 
pattern. This is the Jacobi's integral, $J=E-L\Omega_b$, where $E$ is the energy 
of the particle, $L$ is its angular momentum, and $\Omega_b$ is the pattern 
angular velocity. In such a system no time variations are expected in phase space.
For example, \cite{mq06} showed that no increase in
velocity dispersion with time (disc heating) occurs once the pattern has been 
fully grown. Consequently, in previous such studies
a time-averaging procedure of the particle positions has been employed 
in order to achieve better statistics and minimize computational time  
\citep{fux01,mq07,mq08}.
Unlike in those works, here we look at the time evolution of the system by 
integrating a large number of particles and follow their evolution in phase space.

\subsection{Time evolution of the $u-v$ plane}
\label{sec:uv185}

Figure \ref{fig:uv185} shows the time development of the $u-v$ plane for a
simulation of a stellar disc with Milky Way kinematics perturbed by a steady 
state bar. The pattern speed is $\Omega_b/\Omega_0=1.85$, the maximum sample depth 
is $d_{max}=250$ pc, and the bar orientation is $\phi_0=25^\circ$ (geometry 
as shown in figure \ref{fig:shots}). Shaded contours show the particle number 
density $N$. Each panel is a snapshot of a particular time of the same simulation,
in increments of rotation at $r_0$, up to $t=23$. 

The bar is fully grown in four bar rotations, which corresponds to 
$t\approx2.2$ in the figure. Note that even though the perturbation is stationary, 
structure varies with time. At $t\approx6$ the distribution becomes bimodal,
forming a strong feature at $(u,v)\approx(-10,-20)$.
During the next five rotations this clump moves toward more negative radial 
velocities, becomes weaker and eventually disappears at about $(u,t)=(40,11)$.
This process is repeated approximately every 6.5 rotations, starting at
times $t \approx 6,12$ and $19$. In addition, there is a second feature forming at 
$t\approx7$, which evolves with time to more positive radial velocity $u$, and 
approximately constant $v\approx-10$ km/s. This clump recurs on a slightly 
shorter timescale ($\Delta t\approx$ 5.5) than the one described above, as 
apparent from the two slopes formed by aligning the clumps at $u<0$ and $u>0$, 
for example in the fifth column in figure \ref{fig:uv185}. One should also 
note that the strength of the features diminishes at every period.

Note that the structure in velocity space seen in figure \ref{fig:uv185} is not due to 
our ICs as we initially integrate particles in the axisymmetric potential for 3 Gyr to 
insure the disc is relaxed before the bar is grown. 
The reason for this transient nature of velocity streams in the u-v plane is
an effect caused by the initial response of the disc to the recent formation of 
the bar. To show this we next look at the time evolution of orbits near the 2:1 OLR.

\begin{figure}[t!]
\resizebox{\hsize}{!}{\includegraphics{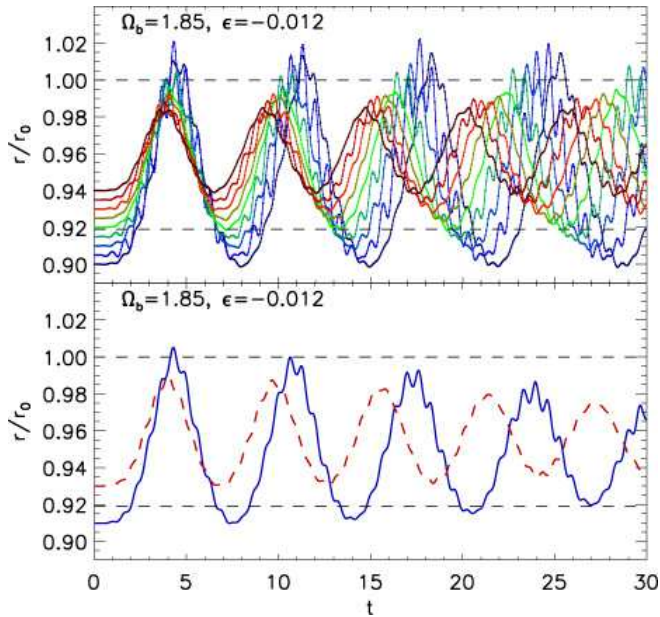}}
\figcaption{
Top: Change in average radius with time for the initially circular rings shown in 
figure \ref{fig:shots}. The solar and OLR radii are indicated by the dashed lines 
at $r=r_0$ and $r\approx0.92r_0$, respectively. 
Bottom: Same as the top panel but for two annuli of radial width $\Delta r=0.02r_0$ 
just inside (blue, solid) and just outside (red, dashed) the 2:1 OLR. 
Red and blue orbits correspond to the streams at $u>0$ and $u<0$ in figure 
\ref{fig:uv185}, respectively.
The amplitude of the oscillations decreases with time as the system relaxes,
similarly to the strength of the velocity streams in figure \ref{fig:uv185}.
\label{fig:rmax}
}
\end{figure}

\subsection{Orbits near the 2:1 OLR of the bar}
\label{sec:shots}

Linear theory predicts that the orientation of orbits shifts across the 2:1 OLR
of the bar, from perpendicular to the bar major axis inside the OLR, to parallel 
with it outside the resonance \citep{bt08}. These are referred to as the 
$x_1(2)$ and $x_1(1)$ orbits, respectively (see, e.g., 
\citealt{quillen03,fux01,dehnen00}). Near the peak of the resonance, both 
types of orbits can exist \citep{cont75,weinberg94}. As suggested by 
\cite{kalnajs91}, for a SN location near the OLR and a bar angle in the range 
$0^\circ<\phi_0<90^\circ$, the closed orbits from either side of the OLR would 
produce two streams; one moving inward ($u>0$) and the other
outward ($u<0$), which could be associated with he Hyades and Sirius stellar 
streams seen in the Hipparcos stellar velocity distribution. Is it possible 
that the two streams we observed in figure \ref{fig:uv185} above
are somehow linked to these orbital families? To try to answer this question,
let us look at how orbits evolve with time.

In figure \ref{fig:shots} we show the time development of nine initially
circular rings near the bar 2:1 OLR, in the bar reference frame. For each ring
particles start at the same initial galactic radius and with a random distributed 
in azimuth. All simulation parameters are the same as in figure \ref{fig:uv185}. 
Rotation is clockwise, thus in the reference frame of the bar
particles move in the counterclockwise direction. The bar is represented
by the ellipse in the center of each panel. The location of the 2:1 ILR and the 
CR are shown as dotted circles and the 2:1 OLR location is indicated by the dashed
circle. The solar neighborhood is shown by a black dot, just outside the 2:1
OLR, for a bar angle of $25^\circ$ (as in figure \ref{fig:uv185}).
The initial radii are such that four rings lie just inside the OLR (blue end) and 
four are just outside the resonance (red end). One ring is situated right on the
resonance (green). 

As the system develops with time, the inner and outer groups of rings tend to 
stay individually aligned and precess at different rates. Since these also
exhibit radial oscillations, in order to see the timescale better
we plotted the average radius of each ring as a 
function of time in figure \ref{fig:rmax}, top panel.
What we would actually see near the Sun is the average contribution from these
orbits. Expecting that the ones just inside the OLR behave differently
than those just outside the resonance, we want to know what is the frequency
of oscillation for each group.  
By considering annuli of finite width $\Delta r=0.02r_0$ (bottom panel of 
figure \ref{fig:rmax}), we can now see that the timescales of precession inside 
and outside the OLR are about 5.5 and 6.5, respectively. These are 
consistent with the recurrence times of the velocity streams in figure 
\ref{fig:uv185}. Thus, we can associate the stream at $u<0$ with a
quasi-periodic family of orbits librating around the $x_1(2)$ closed orbit.
Similarly, the stream at $u>0$ originates from an orbital family librating 
around the $x_1(1)$ close orbit.

The amplitudes of oscillation inside and outside the OLR seen in the
bottom panel of figure \ref{fig:rmax} decay with time; in other words, 
the disc relaxes. This is consistent with the decrease in the fraction of 
particles in the streams in figure \ref{fig:uv185}, as each 
cycle is repeated (see, e.g., $t = 7,13$ and 20). However, stellar streams in the
u-v plane remain quite strong until the end of the simulation at 
$t=23$ LSR rotations. 
This shows that the disc relaxation time is much longer than the age of the
Galaxy. We also expect to see the effect of phase wrapping. For example, at
times $t>17$ in figure \ref{fig:uv185} two steams are apparent at $u<0$ where 
the same orbital family is seen at the end and beginning of its precession around
the $x_1(2)$ orbit. However, in general it appears that phase wrapping effects
are not important, unlike in \cite{minchev09a}.

\begin{figure*}[t!]
\epsscale{1.2}
\plotone{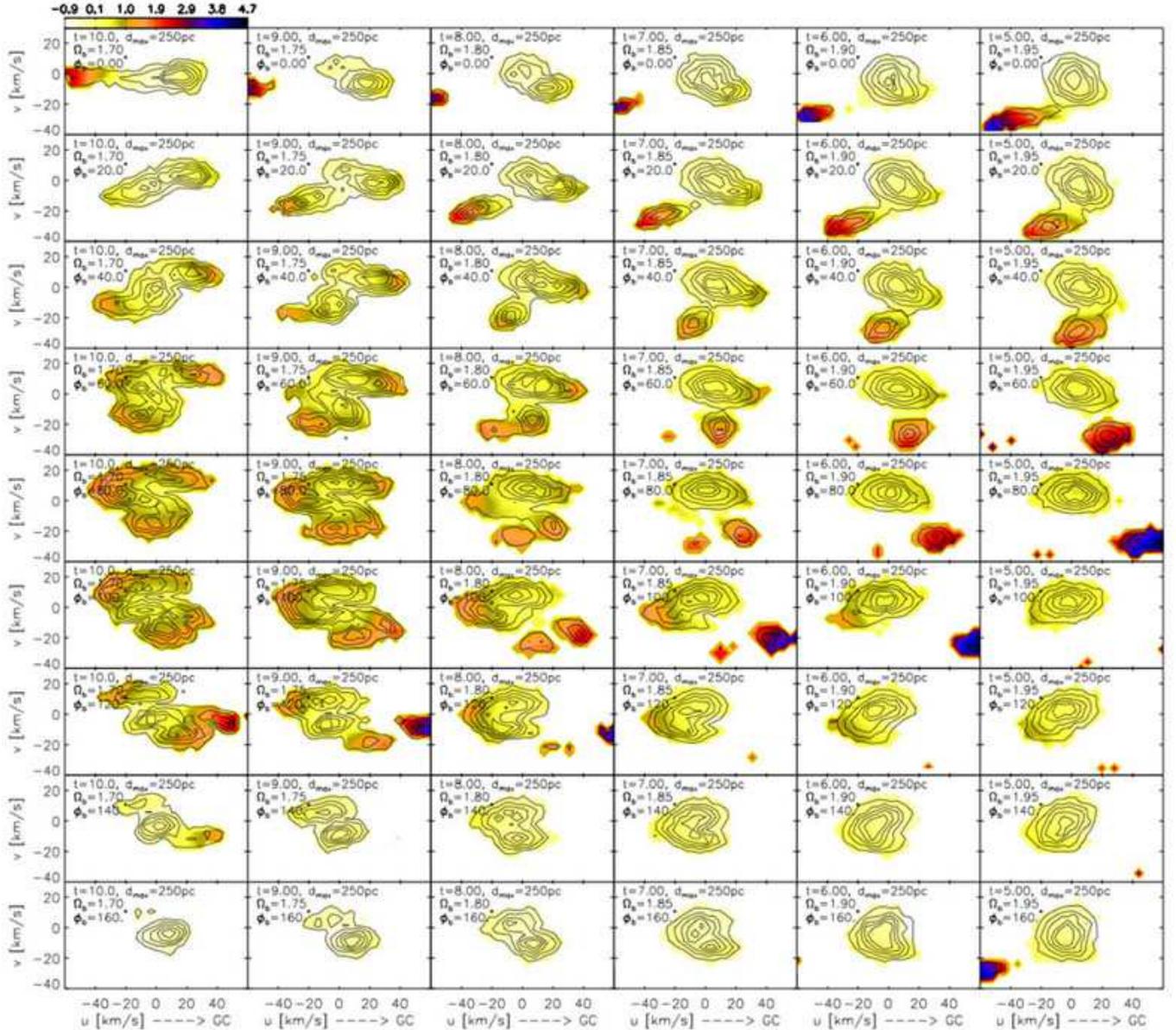}
\figcaption{
Variation in the $u-v$ plane with bar pattern speed and orientation. Contours
show particle number density, while the color levels represent the change in
angular momentum $\Delta L$ for a given location on the u-v plane. Color bar values 
can be converted to (km/s pc) by multiplying by $100 v_0$. Different rows show 
different bar angles from $\phi_0=0^\circ$ to $\phi_0=160^\circ$ in
increments of $20^\circ$. Different columns show changes in bar angular velocity 
for solar circle closer to the 2:1 OLR ($\Omega_b=1.75\Omega_0$) 
to farther away from it ($\Omega_b=2.0\Omega_0$).   
\label{fig:dl}
}
\end{figure*}

\begin{figure}[t!]
\resizebox{\hsize}{!}{\includegraphics{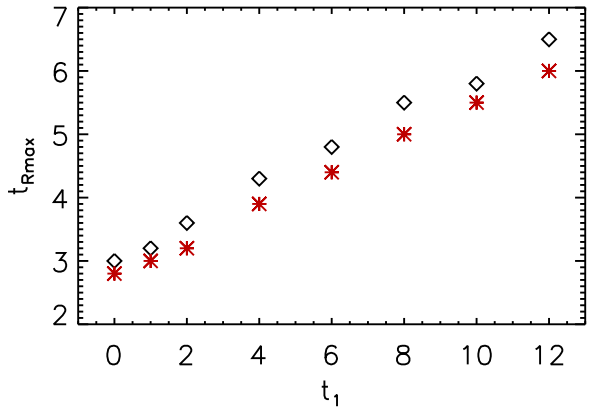}}
\figcaption{
The delay of structure formation as a function of bar formation $t_1$ in
units of bar rotations. The y-axis shows the time at which the functions in 
figure \ref{fig:rmax} peak for the first time. Red and blue symbols correspond
to the orbital families just outside and just inside the bar's OLR. 
\label{fig:tg}
}
\end{figure}

\begin{figure}[t!]
\resizebox{\hsize}{!}{\includegraphics{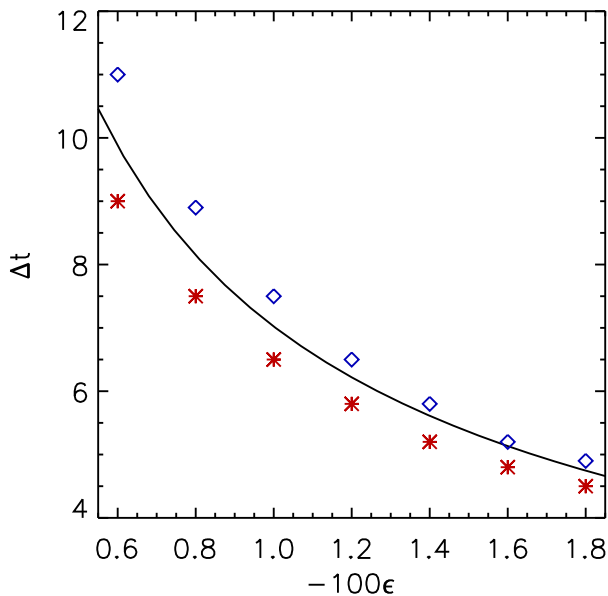}}
\figcaption{
Libration time $\Delta t$ as a function of the bar strength
$\epsilon$. Solid line shows the result from eq. \ref{eq:ham}.
Open blue squares and red star symbols give the precession time for the 
streams just inside and just outside the bar's OLR. The precession time
varies with bar strength as $\epsilon^{-2/3}$.
\label{fig:eps}
}
\end{figure}

\begin{figure*}[t!]
\epsscale{1.2}
\plotone{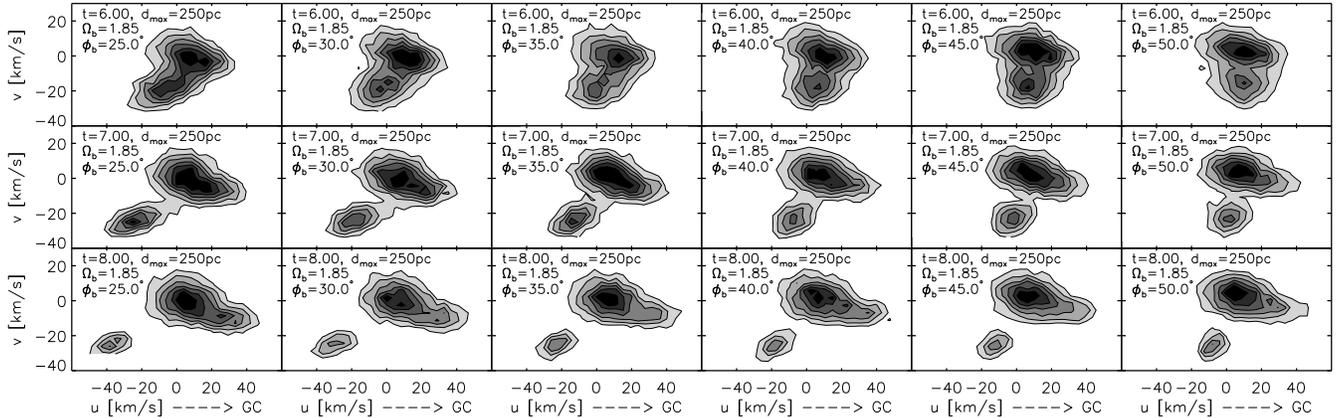}
\figcaption{
Similar to figure \ref{fig:uv185} but showing variations with angle (left to
right) and time (top to bottom) for the same pattern speed
$\Omega_b=1.85\Omega_0$. For this pattern speed, possible matches to the Pleiades
and Coma Berenices streams are achieved for $t=7,8$ and $35^\circ\la\phi_0\la45^\circ$.
\label{fig:185match}
}
\end{figure*}

\begin{figure}[t!]
\resizebox{\hsize}{!}{\includegraphics{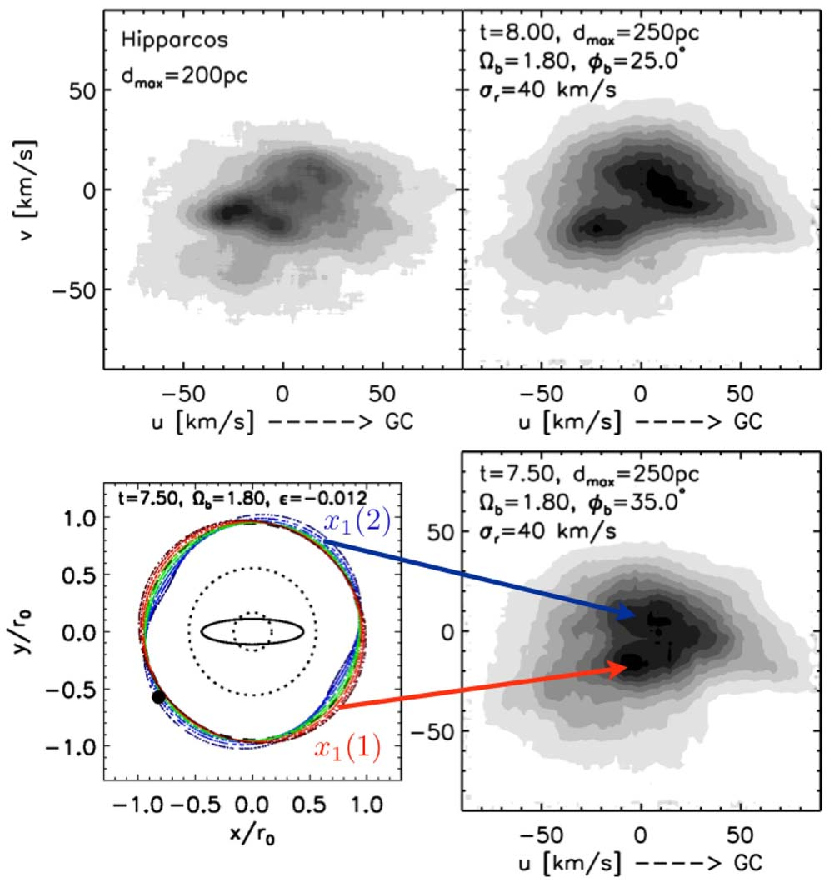}}
\figcaption{
Top left: Hipparcos stellar velocity distribution with the Sun's motion subtracted
(values from \citealt{db98}). Top right: Simulation with initial velocity
dispersion $\sigma_r=40$ km/s and parameters as indicated in the figure. Good match to 
Sirius and Pleiades groups. Bottom right: Small changes in the bar's orientation and 
time since formation, 
provides good match to the Coma Berenices and Pleiades groups. Bottom
left: As in figure \ref{fig:shots}, with simulation parameters as the
corresponding u-v plot on the right. Arrows show the orbital families giving
rise to the streams in the velocity distribution. 
\label{fig:match}
}
\end{figure}

\subsection{Variation with bar pattern speed and orientation}
\label{sec:dl}

Figure \ref{fig:dl} shows simulated SN velocity distributions 
similar to those shown in figure \ref{fig:uv185}. Here however, while the contours
still show the particle number density, the color levels show the change in angular 
momentum $\Delta L=L-L_0$, where $L$ and $L_0$ are the current and initial 
angular momenta. The $\Delta L$ color levels are normalized for all panels, allowing 
one to see the relative change in angular momentum for the full range of parameters 
shown. The color bar values can be converted to (pc km/s) by multiplying by 
$100 v_0$. From top to bottom $\phi_0$ changes from 0 to 
$160^\circ$ in increments of $20^\circ$. Different columns show different pattern 
speeds ranging from $\Omega_b/\Omega_0=1.7-1.95$ (form left to right).
The solar radius is at the bar's OLR in the leftmost column. Particles are 
confined to a region of 250 pc from the Sun.

It is important to realize  that for slower bars (SN closer to the resonance) 
structure develops later than for faster ones (see figure \ref{fig:tg}). 
Consequently, snapshots of the u-v plane at the same time since bar formation
and different angular velocities will show different stages of the precession cycle. 
To approximately account for this, in figure \ref{fig:dl} we show snapshots at
earlier times as the pattern speed increases. For example, when 
$\Omega_b=1.7\Omega_0$ we show a plot at $t=10$, while for $\Omega_b=1.95\Omega_0$ 
$t=5$.

Note the strong variation with both pattern speed and orientation. 
We first look at the variation with angle for the same simulation 
as in figure \ref{fig:uv185}, shown in the forth column of figure \ref{fig:dl}.
The clump at $u<0$ which we associated with the orbital family just inside the OLR 
(see section \ref{sec:shots}), remains at a constant tangential velocity $v$ as
$\phi_0$ changes, but shifts in the positive $u$ direction. It is intriguing 
that at angles $\phi_0>60^\circ$ this stream moves in the positive radial velocities
region. At the same time the stream associated with the $x_1(1)$ closed orbit 
shifts to negative $u$. A similar behavior is observed for all other pattern 
speeds as well. This is easy to understand in view of the precession (in the
reference frame of the bar) and
pulsation of the orbital families just inside and just outside the OLR  
(see figure \ref{fig:shots}). By changing the SN angle, the two families of orbits 
reach the solar vicinity at different orientations, which results in 
sampling different subsets of the orbits in the u-v plane.

Examining a particular row in figure \ref{fig:dl}, we see that as $\Omega_b$ increases 
the streams shift to more negative angular velocities. This is expected if the stars in
the streams come from near the OLR radius, since the distance from the solar
circle to the OLR circle increases with an increase in $\Omega_b$. 
We can see this also in the change in angular momentum shown by the color levels in
figure \ref{fig:dl}. The largest $\Delta L$ occurs for $\Omega_b=1.9\Omega_0$,
meaning that the guiding radius for these stars is smallest. 
Note that since the expected 
bar angle is in the range $10<\phi_b<60$, possible matches to the observed
velocity field will always have $u<0$ for the $x_1(2)$ family and $u>0$ for the
$x_1(1)$ one.

As apparent form examining the fifth row in figure \ref{fig:dl} we 
see that beyond $\Omega_b=1.85\Omega_0$ phase wrapping is more efficient near the 
resonance and becomes unimportant away from it.

\subsection{Variation with bar growth time}

In this paper in all figures we grow the bar in four bar rotation.
In order to see how our results would change with a change in the bar growth 
time $t_1$, we ran simulations with different values in the range $0<t_1<12$ 
in units of bar rotation. We plot the results in figure \ref{fig:tg}. The y-axis 
shows $t_{rmax}$, the time for which the functions in figure
\ref{fig:rmax} peak for the first time. The x-axis shows the bar growth time
used for that particular simulation.
Red and blue symbols correspond to the orbital families just outside and just 
inside the bar's OLR. The dependence is almost linear with
$t_{rmax}\approx1/4t_1+3$. The relative phases of the functions in figure
\ref{fig:rmax} remain the same for all values of $t_1$, as do all our other results. 
We conclude that while this delay in structure formation as $t_1$ increases
would introduce uncertainty in our prediction for the
bar formation/evolution time, the general effect of bar growth time in
unimportant.

\subsection{Variation with bar strength}

Our default value for the amplitude of the bar perturbation is $\epsilon=-0.012$. 
We would like to know how changing this will affect our results. We first derive an 
analytical expression for the orbital libration timescale $\Delta t$, as a function 
of bar strength $\epsilon$. Next we compare that to our results from simulations.

\cite{cont75,cont88} showed that the dynamics of stars confined to the Galactic
plane in the case of an axisymmetric potential can be described by a Hamiltonian  
written in a third order post-epicyclic approximation.
Taking it further, \cite{quillen03} considered the dynamics of stars that are
affected by perturbations from both spiral structure and the Milky Way bar. She
constructed a simple one dimensional Hamiltonian model for the strongest
resonances in the epicyclic action angle variables. We can use her equation 22 
to describe the effect of the bar only, by neglecting the term related to spiral
structure (last term on the right hand side).  
In a slightly different notation we can write
\begin{equation}
H = a p^2 + \delta p + \beta p^{1/2} \cos{\phi},
\end{equation}
where $p$ is the action variable related to eccentricity
and $\phi$ is the resonant angle. The coefficient $a=-5.7r_0^{-2}$ (eq. 18 in 
\citealt{quillen03}) is constant at the solar radius and 
$\delta=\kappa+2(\Omega-\Omega_b)$ sets the distance from resonance. For a flat 
rotation curve $\beta\approx0.086\sqrt{2/\kappa}(A_f/r_0)(r_b/r_0)^3$ and is related to 
the bar amplitude 
$\epsilon$ through $A_f = (\epsilon v^2_0/3)(r_0/r_b)^3$ (eq. 7 in
\citealt{dehnen00}), where $r_b\approx0.43r_0$ is the radius at which the bar
ends (for $\Omega_b=1.85\Omega_0$). By dimensional analysis from $a$ and $\beta$ 
we can derive a timescale for the resonance:
\begin{equation}
\label{eq:ham}
\Delta t \approx (2\pi)^{-1}a^{-1/3} \beta^{-2/3},
\end{equation}
where we have divided by $2\pi$ to put the expression in units of LSR rotations.
This is the libration time around the fixed points associated with the $x_1(1)$ and 
$x_1(2)$ orbits. Note that the functional
dependence of libration time on bar strength is 
\begin{equation}
\Delta t\sim\epsilon^{-2/3}.
\end{equation}

To compare to this analytical result we performed simulations with different 
perturbation strengths, in addition to
the default value of $\epsilon=-0.012$. 
In figure \ref{fig:eps} the open blue squares
and the red star symbols show the precession time for the streams just inside
and just outside the OLR for a bar pattern speed as in figures \ref{fig:uv185}-
\ref{fig:rmax} and different bar perturbation amplitude in the range 
$-0.006\leqslant\epsilon\leqslant-0.018$. The solid line plots equation \ref{eq:ham} 
scaled by a factor of 0.38 (estimates by \citealt{quillen03} are accurate to within an 
order of magnitude). We see excellent agreement in the functional behavior between 
our numerical and analytical results.
Simulations with stronger bars result in shorter libration time. 
Features in the u-v plane remained of similar
strength to those seen in figure \ref{fig:uv185}.

\section{Constraining the bar}

As we showed in section \ref{sec:shots}, for a pattern speed consistent with 
expectation for the MW bar, we can associate two streams in velocity space
with the quasi-periodic families of orbits librating around the $x_1(2)$
and $x_1(1)$ closed orbits. In section \ref{sec:dl} we explored the variation
of structure in the u-v plane with angular velocity and bar angle for a
particular time. Now we would like to see how we can use this information to
put constraints on bar parameters, such as pattern speed, orientation
and formation time. 

For a given time, the tangential velocity $v$ of resonant features in the u-v 
plane is set by the 
bar pattern speed as evident from figure \ref{fig:dl}. On the other hand, the
radial velocity $u$, is set by the bar's orientation. Thus, assuming some 
features at low velocities in the Hipparcos velocity distribution are of resonant 
origin, we can match stream positions in the u-v plane and estimate $\Omega_b$ and 
$\phi_0$. In addition, structure varies with time due to the libration of the 
quasi-periodic orbits around the fixed points; this allows us to constrain the 
bar formation time.
We need to compare our synthetic velocity distributions to the Hipparcos u-v
plane corrected for the motion of the Sun. As recently discussed by
\cite{mcmillan09}, the velocity of the Sun with respect to the LSR, velocity of
the LSR and the Sun's Galactocentric radius may be quite different than the
currently accepted values. However, the LSR angular rotation rate $\Omega_0=v_0/r_0$
is found to be well constrained in the range 29.9-31.6 km/s/kpc. 
$\Omega_0$ sets the relative position of streams in the u-v plane. Therefore,
as we consider possible matches to the observed velocity distribution we need to 
be most concerned with the relative orientation of streams. In this paper
we use the values $v_0=240$ km/s and $r_0=8$ kpc resulting in $\Omega_0=30$
km/s/kpc.

In figure \ref{fig:185match} we show how structure in the u-v plane changes with
bar angle and time, for the simulation in figure 
\ref{fig:uv185} ($\Omega_b=1.85\Omega_0$). While the pattern speed is kept fixed
everywhere, different columns show variation with angle in the range 
$25^\circ<\phi_0<50^\circ$ in increments of $5^\circ$. Rows show
changes with time in units of rotations at $r_0$ for $t=6,7,$ and 8. 
The clump at negative $u$ can be associated with the Pleiades moving group,
while the one at $u>0$ at the beginning of its formation is consistent
with the Coma Berenices stream.For this pattern speed, good matches are
achieved for $t=7,8$ and $35^\circ\la\phi_0\la45^\circ$.
Since structure in the u-v plane is recurrent about every six LSR rotations
(see figure \ref{fig:uv185}), one would expect later
times to be equally consistent. However, we require a sufficiently strong
signal in order to account for the fraction of stars in the Pleiades, thus we
reject later times.

In figure \ref{fig:match}
we compare the observed velocity distribution to our simulations.
We used the catalog from \cite{holmberg09}
to plot the u-v distribution of Hipparcos stars in the top left panel of figure
\ref{fig:match}. To subtract the Solar motion we used the values by \cite{db98}. 
In the top right panel we show a simulation with parameters as indicated in the
figure. Note that here we start with a radial velocity dispersion $\sigma_u=40$
km/s in order to populate the Hercules stream at $v\approx=40$ km/s. The two
clumps in the top right panel are consistent with the Pleiades and Sirius
groups. For a $10^\circ$ change in bar orientation and half a rotation at $r_0$
we can associate these groups of stars with the Pleiades and Coma Berenices
streams (bottom right). Finally, the bottom left panel in figure \ref{fig:match}
shows the orbital families giving rise to the streams in the u-v plane.

Similar inspection of the u-v plane for pattern speeds in the range 
$1.7\leqslant\Omega_b/\Omega_0\leqslant2.0$ reveal that, depending on the bar 
parameters and time since bar formation, we can obtain a match to either the Coma 
Berenices or Sirius moving groups in addition to the Pleiades stream. We conclude 
to be consistent with structure in the observed u-v velocity distribution our  
model requires the bar angle to be in the range 
$30^\circ\leqslant\phi_0\leqslant45^\circ$ and its pattern speed to be 
$\Omega_b/\Omega_0=1.82\pm0.07$. This is in very good agreement with previous 
estimates of these bar parameters. For example, by a quantitative comparison 
of the observed with the simulated position of the Hercules stream,
\cite{dehnen00} deduced the MW bar pattern speed to be
$\Omega_b/\Omega_0=1.85\pm0.15$ and the bar angle
$10^\circ\leqslant\phi_0\leqslant70^\circ$. By relating the dynamical effect of the 
bar to the derivatives of the velocities via the Oort constant C, an additional 
constraint on bar parameters was provided by \cite{mnq07}, where they estimated
$\Omega_{\rm b}/\Omega_0=1.87\pm0.04$ and
$20^\circ\leqslant\phi_0\leqslant45^\circ$. Our result is also in agreement
with the estimate by \cite{debattista02}, based on OH/IR star kinematics.

To account for the strength of the Pleiades, we 
find that the bar formation time has to be between 7 and 9 rotations at $r_0$, 
which corresponds to 1.75-2.25 Gyr.

\subsection{Relating to other work}

Previous works have also reported features (or not reported but are identifiable 
in their figures) in their barred disc simulations, which we can now explain as 
the initial response of the disc to bar formation. 

One example is the work by \cite{bagley08} who investigated the 
effect of a central bar on disc morphology. 
In their figure 3 they present number density plots of the time evolution of
a barred galactic disc, similar to our figure \ref{fig:shots}, but showing the 
extent of the full disc. In addition to the asymmetries in the $R2$ ring 
(associated with the $x_1(2)$ orbit) reported by the authors, it is easy to identify
also a second ring at smaller radius and different orientation relative to the 
outer one ($x_1(1)$ orbit). Even at the last three time outputs (figure ends at 
$t=12.5$ in our units) where the outer ring appears stable, close inspection of 
the region between the two rings reveals misaligned features similar to those 
seen in our figure \ref{fig:shots}. 

A second example is the work by \cite{fux01}, where the author investigated the
effect of the bar on the u-v plane. The second row of his figure 12 presents
simulated local velocity distributions for a ratio of solar to OLR radius 
$r_0/r_{OLR}=1.1$ (or equivalently, $\Omega_b=1.87\Omega_0$) and different 
bar angles. Due to the time-averaging procedure (from 13 to 19 rotations in 
our units) used by Fux we can compare a particular u-v plot from his figure 
to the time average of all panels in our figure 
\ref{fig:uv185}. For an angle of $30^\circ$ (second row, second column in 
figure 12 of \citealt{fux01}), we can see two features stretching 
from the central clump to radial velocities of $u\approx\pm0.2v_0\approx\pm40$ 
km/s. Just as expected from our results, the feature at positive $u$
(corresponding to negative $u$ in our plots) reaches more negative tangential
velocities than the one moving away from the galactic center. 
Similar structure is apparent for other pattern speeds and angles as well. 
In contrast, in figure 13 of the same paper, where the velocity distributions 
are time averaged between 55 and 65 bar rotations, the features we described 
above are almost not apparent, as the disc is more relaxed at later times. 

A third example can be found in \cite{antoja09}, who examined the effect of the 
Galactic bar and spiral structure. Similarly to the present study, they did not 
time-average over particle orbits. Although no time development of the u-v plane was 
presented, it is easy to identify some structure at low velocities in their Fig. 1(f). 
Note also that their discs were unrelaxed, similar to 
those of \cite{minchev09a}, which is probably the reason for the extended feature
(in the u direction) at $(u,v)\approx(-20,-30)$ km/s.

\section{Discussion and conclusions}
\label{sec:concl}

In this paper we have examined the effect of a central bar on the
structure in the stellar velocity distribution in the solar neighborhood.
Unlike in previous works, here we do not time average over particle positions
but follow the time evolution of the system. We find that in addition to the
central clump in the u-v plane, two strong features
appear after the formation of the bar, which shift with time toward
more positive and more negative radial velocities (see figure \ref{fig:uv185}). 
This is surprising since a system subjected to a steady state periodic 
perturbation must not evolve with time due to the presence of an isolating 
integral of motion (the Jacobi's integral). We explain these results as the
initial response of the disc following the bar formation. We associate the 
streams in the u-v plane with two quasi-periodic orbital families precessing 
around the $x_1(1)$ and $x_1(2)$ closed orbits near the 2:1 OLR of the bar. 
The timescale of this libration is dependent on the bar's strength as 
$\Delta t\sim\epsilon^{-2/3}$, consistent with a simple Hamiltonian model for
the resonance (eq. \ref{eq:ham}). We find that phase wrapping timescales are 
longer than libration timescales thus the disc takes a long time to relax near
the 2:1 OLR of the bar.

We have shown for the first time that the Milky Way bar could be responsible
for low-velocity structure in the solar neighborhood velocity field. 
The stream caused by the orbital family just inside the OLR of the
bar (figure \ref{fig:match}) can be associated with the Pleiades moving group
seen in the Hipparcos velocity distribution. Similarly, the orbital family just
outside the OLR can account for either the Coma Berenices or the Sirius 
velocity stream, depending on the bar parameters and time since its formation.
We note that the position of the Hercules stream (figure \ref{fig:match}) does not
change with time as in the case of the lower velocity streams. 
This model requires a bar pattern speed of $\Omega_b/\Omega_0=1.82\pm0.07$ and 
a bar orientation $30^\circ\leqslant\phi_0\leqslant45^\circ$.
Since the process is recurrent, we achieve a good match to the velocity
distribution about every six rotations of the LSR. However, in order to be
consistent with the fraction of stars in the Pleiades velocity stream, we
favor a bar formation time 1.75-2.25 Gyr ago. This is consistent with the
observational results by \cite{cole02}, who estimated that the Milky
Way bar is likely to have formed more recently than 3 Gyr ago.
In addition, the recent work by \cite{minchev09a} in which the authors considered
the effect of a Milky Way merger on the kinematics of stars in the SN,
estimated that the Galactic disc was strongly perturbed about 2 Gyr ago.
Such a merger could have triggered bar formation.

With the upcoming data from the Gaia mission,it will become possible to see how 
the positions 
of the streams we here attribute to the effect of the Milky Way bar, change with 
Galactic azimuth and Galactocentric distance, and thus test our model. Furthermore,
it might be possible to identify the $x_1(1)$ and $x_1(2)$ orbital families in 
pencil-beam surveys such as ARGOS, BRAVA, and APOGEE by observing, for example, 
in the low absorption window at a Galactic longitude of $\sim290^\circ$ 
\citep{carraro09}, and identifying structure in the way suggested by \cite{mq08}. 
 
This work confirms that resonant features in velocity space can provide tight 
constraints on nonaxisymmetric disc structure parameters, as shown by 
\cite{qm05} for the case of spiral density waves. We expect that spiral
structure can also cause precession of resonant orbital families in the reference 
frame of the perturber as seen here, and thus provide an insight on spiral structure 
formation. Preliminary results from spiral density wave simulations show that this 
is indeed true (Minchev et al. 2010 , In preparation). However, due to the more narrow
resonant widths compared to the case of the bar, this effect might not be as
important unless the solar radius is almost at the exact resonance with spirals.

Decreasing the pattern speed moves the bar's resonances outwards, causing orbits 
to be captured into resonance 
(e.g. \citealt{romero06,bagley08}). Therefore, it is likely that the features in the u-v 
plane will remain strong for longer periods of time than what is inferred from 
figure \ref{fig:uv185}. Consequently, the model presented here might also be 
consistent with an old bar, provided the bar pattern speed is changing. Further 
investigation exploring this problem is necessary. An additional perturbation from 
the Milky Way spiral structure will change the dynamics as well \citep{qm05,desimone04,
antoja09}. 
Models incorporating both spiral and bar structure, as the one considered by 
\cite{chakrabarty07} but without suppressing the effects of the initial disc response, 
are needed as well. We have recently shown that such a configuration provides a very 
powerful stellar radial migration mechanism \citep{mf09}, however, the effect 
on the u-v plane is yet to be explored. 
We note that the parameter space in the case of two perturbations becomes enormous,
considering the timescales of each perturbing agent associated with the initial 
response of the disc, in addition to the relative orientation and individual time
evolution of spirals and bar. Thus a thorough investigation of the effect
of spiral structure (including transient spirals) and transient bars is required
before a superposition of these is considered. Even more complications to the
disc dynamics will be added if the effect of an orbiting satellite is considered 
\citep{quillen09}.

The model in this paper argues against a common dynamical origin for the Hyades and 
Pleiades moving groups, in agreement with the results of analysis of the
Hipparcos catalogue \citep{famaey08,bovy10}. This raises the question of whether it is 
the MW spiral structure responsible for the Hyades as expected from the work by 
\cite{qm05}. A work investigating this problem is underway (Minchev et al. 2010, 
In preparation).

\acknowledgements
We thank Alice Quillen for valuable suggestions that have greatly
improved the manuscript. 
Support for this work was provided by ANR and RAVE.


\end{document}